\documentstyle{mn}

\input epsf

\begin{document}
\journal{Sussex preprint SUSSEX-AST 97/12-1, SUSX-TH-97-023, 
astro-ph/9712028}
\title[Cosmological parameter estimation and the spectral index from 
inflation]{Cosmological parameter estimation and the spectral index from 
inflation}
\author[E.~J.~Copeland, I.~J.~Grivell and A.~R.~Liddle]{Edmund 
J.~Copeland,$^1$ Ian J.~Grivell$^2$ and Andrew R.~Liddle$^2$\\
$^1$Centre for Theoretical Physics, University of Sussex, Falmer, Brighton 
BN1 9QH\\
$^2$Astronomy Centre, University of Sussex, Falmer, Brighton BN1 9QH}
\maketitle
\begin{abstract}
Accurate estimation of cosmological parameters from microwave
background anisotropies requires high-accuracy understanding of the
cosmological model. Normally, a power-law spectrum of density
perturbations is assumed, in which case the spectral index $n$ can be
measured to around $\pm 0.004$ using microwave anisotropy satellites
such as {\it MAP} and {\it Planck}. However, inflationary models
generically predict that the spectral index $n$ of the density
perturbation spectrum will be scale-dependent. We carry out a detailed
investigation of the measurability of this scale dependence by {\it Planck},
including the influence of polarization on the parameter
estimation. We also estimate the increase in the uncertainty in all
other parameters if the scale dependence has to be included.  This
increase applies even if the scale dependence is too small to be
measured unless it is {\em assumed} absent, but is shown to be a small
effect.  We study the implications for inflation models, beginning
with a brief examination of the generic slow-roll inflation situation,
and then move to a detailed examination of a recently-devised hybrid
inflation model for which the scale dependence of $n$ may be
observable.
\end{abstract}
\section{Introduction}

One of the most important legacies of the {\it COBE} satellite for the 
inflationary cosmology was its emphasis that observations had reached a 
quality where the inflationary prediction for the density perturbations 
could no longer be taken to be the scale-invariant Harrison--Zel'dovich 
spectrum. That paradigm was quickly replaced by a new one, where the 
spectrum is approximated by a power-law in wavenumber, with spectral index 
$n$ giving the scale dependence. Different inflation models predict 
different values for this spectral index, and the high accuracy with which 
$n$ can potentially be observed (Bond, Efstathiou \& Tegmark 1997; 
Zaldarriaga, Spergel \& Seljak 1997) promises very strong 
discrimination between different inflation models \cite{LL,L96}.

This raises the question of whether future observations, in particular
high-accuracy microwave background anisotropy measurements by
satellites such as {\it MAP} and {\it Planck}, might be of such
stunning accuracy that even the power-law approximation may prove
inadequate. This has long been known to be the case for `designer'
models of inflation \cite{KL,SBB,HBKP,HB}, where one tunes sharp
features into the power spectrum by careful placement of strong
features in the potential of the inflation field, at just the point
where perturbations on cosmologically observable scales are being
created. Such models have long been regarded as rather unnatural, but
retain interest as they are eminently testable by combinations of
microwave background and large-scale structure observations.

A more pertinent question concerns whether the breaking of the
power-law might be detectable in the types of models regarded as
theoretically most appealing. Such models are of the slow-roll type
(see Liddle \& Lyth 1993 for a review). Several papers had pointed out
that the spectral index would not be constant in typical models
(Barrow \& Liddle 1993; Copeland et al.~1993, 1994a), and an analysis
of a range of slow-roll models was given by Kosowsky \& Turner
\shortcite{KT} who wrote down a general formula for the scale
dependence using the slow-roll expansion.  Since their paper, however,
theoretical prejudice has moved towards a subset of slow-roll
inflation models, the so-called hybrid inflation models
\cite{L1,L2,CLLSW,L96}. These models have the disappointing feature 
observationally that any gravitational wave contribution to the microwave 
anisotropies is expected to be negligible \cite{L97}, but at least in some 
cases this appears to be compensated by a possible detectability of the 
scale dependence of the spectral index.

An important question is the influence of the possible scale
dependence on estimates of all other cosmological parameters, through
the extra degeneracies introduced. We shall show that the cosmological
parameters are only mildly affected.

\section{Observational capabilities}

\label{s:ian}

One of the crucial roles of microwave anisotropy experiments is to
estimate cosmological parameters. These can be divided into two
types. The cosmological parameters, such as the Hubble parameter, the
density parameter and parameters describing the dark matter, give the
evolution of the background space-time. The second set of parameters
describe the density perturbations which lead to the anisotropies. The
simplest assumption, which coincides with the prediction of most
inflation models and which we shall adopt throughout, is that the
perturbations are adiabatic and Gaussian-distributed, in which case
they are completely specified by the form of the density perturbation
spectrum.

It goes without saying that the more parameters we believe are needed
to describe a data-set, the worse determined they will be, so the
input power spectrum plays a crucial role in the parameter
estimation. Assuming at least that the power spectrum is suitably
slowly varying, the best way to proceed (see e.g.~Lidsey et al.~1997)
is to expand its logarithm as a Taylor series in $\ln k$ (where $k$ is
the comoving wavenumber) about some wavenumber $k_0$, and truncate
after some number of terms, i.e.
\begin{eqnarray}
\label{psexp}
\ln {\cal P}_{\cal R}(k) & = & \ln {\cal P}_{\cal R}(k_0) + (n-1) \ln
	\frac{k}{k_0} \\
 & & \hspace*{-1cm} + \frac{1}{2} \left. \frac{dn}{d \ln k} \right|_{k_0} 
 	\, \ln^2 \frac{k}{k_0} + \frac{1}{6} \left. 
 	\frac{d^2n}{d (\ln k)^2}
	\right|_{k_0} \, \ln^3 \frac{k}{k_0} +\cdots \nonumber
\end{eqnarray}
Here ${\cal P}_{\cal R}(k)$ is the spectrum of the curvature
perturbation, defined as in Liddle \& Lyth \shortcite{LL}, which is
scale-independent for a Harrison--Zel'dovich spectrum, which
corresponds to $n=1$. Obviously $k_0$ is best chosen near the center
of the range probed by the observations in question, but we will be
more precise about this below.

The Harrison--Zel'dovich spectrum amounts to truncating this expansion
after one term, and the power-law approximation to truncating it after
the second term. We shall investigate the possible inclusion of the third
and fourth terms.

Whether or not the power-law approximation is accurate clearly depends
on the range of scales probed, measured by the size $\ln k/k_0$
reaches for a given data set. For {\it Planck}, which probes a range
of multipoles from $\ell = 2$ to about $2000$, this term reaches about
$\pm 3$ on either side of the central value. It is known that {\it
Planck} can measure the spectral index $n$ of a perfect power-law
spectrum to an accuracy of better than $\pm 0.01$; however this does
not directly lead to an estimate of the uncertainty in $dn/d\ln k$
since this uncertainty also depends on the sensitivity of the given
experiment to the power spectrum as a function of scale. Instead, we
determine the errors on the first two derivatives of $n$ by including
them in the set of parameters to be determined from the data and using
the Fisher matrix, as we now describe.

In principle, a brute-force maximum likelihood approach would give the
best estimates of a set of cosmological parameters from a CMB
anisotropy data set, in the sense that the expected variances of the
estimates are minimized. In practice, such an approach would not be
feasible for the quantity of data expected from the next generation of
experiments (for example, satellite experiments will produce CMB maps
with at least $10^6$ pixels). However, recent work on data analysis
techniques suggests that near-optimal estimates of parameters could be
achieved using computationally feasible algorithms \cite{T,Betal2}. In
any case, it is useful to consider what the smallest possible error
bars on the parameter estimates will be and these can be calculated in
a straightforward way using the Fisher information matrix,
$\alpha_{ij}$ \cite{TTH}. The covariance matrix for the parameter
estimates is then the inverse of this matrix, and in particular the
standard error for the estimate of parameter $s_i$ is
$\sqrt{(\alpha^{-1})_{ii}}$. This approach was used by Jungman et al.
\shortcite{Jetal1,Jetal2} to determine the accuracy with which
cosmological parameters could be determined from observations of
temperature anisotropies, making use of a semi-analytical method to
calculate the required angular power spectra and allowing for a
variation in the spectral index with scale. Bond et
al. \shortcite{Betal} repeated their analysis, calculating the angular
power spectra with a numerical Boltzmann code for greater accuracy,
but did not consider deviations from a power-law perturbation
spectrum. The improvement in parameter estimates when polarization
information is included was investigated by Zaldarriaga et
al.~\shortcite{Zetal}. We now generalize their results by including
$dn/d\ln k$ and $d^2n/d (\ln k)^2$ in the set of parameters to be
determined.

\begin{table*} 
\begin{minipage}{13cm}
\centering
\caption{\label{errors_tab} Estimated parameter errors (one-sigma) for the 
Standard CDM model, under different assumptions for experimental 
configuration and generality of the underlying cosmological model.}  
\begin{tabular}{lllllllll} 
Parameter &\hspace{0.5cm}& \multicolumn{3}{c}{Planck HFI} & \hspace{0.5cm} &
\multicolumn{3}{c}{Planck 140 GHz channel} \\ 
 & &\multicolumn{3}{c}{no polarization}& & \multicolumn{3}{c}{with
polarization} \\ \hline
$\delta \langle {\cal C}_{\ell} \rangle_{{\rm B}}^{1/2}/\langle {\cal 
C}_{\ell}  
 \rangle_{{\rm B}}^{1/2}$ & &
$  0.01 $ & $  0.01 $ & $ 0.01 $ & & $  0.005$ & $  0.008$ & $ 0.01$ \\
$\delta \Omega_{{\rm b}} h^2 /\Omega_{{\rm b}} h^2$ & &
$ 0.006 $ & $  0.01 $ & $ 0.01 $ & & $  0.007$ & $  0.008$ & $ 0.009$ \\
$\delta \Omega_{{\rm nr}} h^2 /h^2$ & &
$  0.02 $ & $  0.02 $ & $ 0.03 $ & & $   0.02$ & $   0.02$ & $ 0.02$ \\
$\delta \Omega_{{\rm vac}} h^2 /h^2$ & &
$  0.06 $ & $  0.07 $ & $ 0.08 $ & & $   0.04$ & $   0.04$ & $ 0.05$ \\
$\delta \tau_{{\rm C}}$ & &
$   0.12 $ & $   0.15 $ & $ 0.15  $ & & $ 0.002$ & $ 0.002$ & $ 0.002$\\ 
\\
$\delta n$ & &
$ 0.006 $ & $   0.007 $ & $ 0.007  $ & & $  0.004$ & $   0.004$ & $ 0.006$ \\
$\delta r_{{\rm ts}}$ & &
$   0.12 $ & $   0.14 $ & $ 0.14  $ & & $   0.04$ & $   0.04$ & $ 0.04$ \\
$dn/d\ln k$ & &
$      - $ & $  0.02 $ & $ 0.02 $ & & $       -$ & $  0.009$ & $0.01$ \\
$d^{2}n/d(\ln k)^2$ & &
$      - $ & $      - $ & $ 0.02$ & & $       -$ & $       -$ & $0.02$ \\
\hline
\end{tabular}
\end{minipage}
\end{table*}

When both temperature and polarization anisotropy
observations are available the Fisher information matrix is given by
\cite{Zetal} 
\begin{equation} \label{fisher}
\alpha_{ij} = \sum_l \sum_{X,Y} \frac{\partial C_{Xl}}{\partial s_i}
{\rm Cov}^{-1}(\hat{C}_{Xl},\hat{C}_{Yl})\frac {\partial C_{Yl}}
{\partial s_j}\, ,
\end{equation}
where $s_i$ are the parameters to be estimated and $C_{Xl}$ are the
angular power spectra, with $X$ and $Y$ standing for $T$
(temperature), $E$ (even parity polarization), $B$ (odd parity
polarization) and $C$ ($E$ and $T$ cross correlation)
anisotropies. ${\rm Cov}^{-1}$ is the inverse of the covariance matrix
between the estimators of the power spectra. The diagonal elements of
this matrix are 
\begin{eqnarray}
{\rm Cov}(\hat{C}^2_{Tl}) &= &\frac{2}{(2l+1)f_{\rm sky}}(C_{Tl} +
	w_T^{-1} e^{l^2 \sigma_b^2})^2 \\
{\rm Cov}(\hat{C}^2_{El}) &= &\frac{2}{(2l+1)f_{\rm sky}}(C_{El} +
	w_P^{-1} e^{l^2 \sigma_b^2})^2 \\
{\rm Cov}(\hat{C}^2_{Bl}) &= &\frac{2}{(2l+1)f_{\rm sky}}(C_{Bl} +
	w_P^{-1} e^{l^2 \sigma_b^2})^2 \\
{\rm Cov}(\hat{C}^2_{Cl}) &= &\frac{1}{(2l+1)f_{\rm sky}} \times 
	\\
 & & \hspace*{-0.5cm} [C_{Cl}^2 +(C_{Tl} + w_T^{-1} e^{l^2 \sigma_b^2}) 
	(C_{El} + w_P^{-1} e^{l^2 \sigma_b^2})], \nonumber
\end{eqnarray}
and the non-zero off-diagonal elements are
\begin{eqnarray}
{\rm Cov}(\hat{C}_{Tl}\hat{C}_{El}) &= &\frac{2}{(2l+1)f_{\rm sky}} 
	C^2_{Cl} \\
{\rm Cov}(\hat{C}_{Tl}\hat{C}_{Cl}) &= &\frac{2}{(2l+1)f_{\rm sky}}
	C_{Cl}(C_{Tl} + w_T^{-1} e^{l^2 \sigma_b^2}) \\
{\rm Cov}(\hat{C}_{El}\hat{C}_{Cl}) &= &\frac{2}{(2l+1)f_{\rm sky}}
	C_{Cl}(C_{El} + w_P^{-1} e^{l^2 \sigma_b^2})\,.
\end{eqnarray}
Here $w_T$ and $w_P$ characterize the noise in temperature and
polarization measurements respectively, $f_{\rm sky}$ is the fraction
of the sky sampled and $\sigma_b$ is the beam width for the
experiment. We calculate parameter uncertainties for two different
specifications for the {\it Planck} satellite -- one for four channels
of the High Frequency Instrument (HFI) with no polarization
capability (using the experimental parameters of Bond et al.~1997) and
one for the 140 GHz channel of the HFI with polarizers (using the
parameters of Zaldarriaga et al.~1997). In both 
cases we take the usable sky area as $f_{\rm sky}=0.65$.

In addition to the specifications of the experiment being considered,
the estimated parameter errors also depend on the underlying
cosmological model from which the data is drawn and on the choice of
parameters to be determined from the data. For the latter we adopt a
subset of the variables used by Bond et al. \shortcite{Betal} along
with the derivatives of the scalar spectral index mentioned above. The
densities of the various types of matter are specified by the
parameters $\omega_j \equiv \Omega_j h^2$, where $h$ is the Hubble
parameter in units of \mbox{100 km s$^{-1}$ Mpc$^{-1}$}, and $j$=b,
nr, vac refers to baryons, non-relativistic matter (baryons and cold
dark matter combined) and the energy density associated with a
cosmological constant, respectively. We assume there is no hot dark
matter component and also that the Universe is flat, so the Hubble
parameter is given by $h^2= \omega_{{\rm nr}} + \omega_{{\rm
vac}}$. The ionization history is characterized by the Compton optical
depth $\tau_{{\rm C}}$ from the redshift of reionization to the
present, assuming complete reionization throughout that period. The
normalization of the power spectrum is given in terms of the
band-power corresponding to the filter for the given experiment,
$\langle {\cal C}_{\ell} \rangle _{{\rm B}}^{1/2}$ \cite{Bond}.
Finally, we consider the inflationary parameters: $r_{{\rm ts}}$ which is
the ratio of the tensor to scalar quadrupole moments, and the scalar
spectral index $n$ and its derivatives. They are discussed more fully
in the next section. For the underlying cosmology we take a Standard
Cold Dark Matter (SCDM) model, with the following parameters; $h=0.5$,
$\Omega_{{\rm b}}=0.05$, $\Omega_{{\rm CDM}}=0.95$, $\Omega_{{\rm
vac}}=0$, $\tau_{{\rm C}}=0$, $r_{{\rm ts}}= 0$, $n=1$, $dn/d\ln k=
d^2 n/ d(\ln k)^2= 0$.

The calculation of the Fisher matrix Eq.~(\ref{fisher}) requires the
derivatives of the angular power spectra with respect to each
parameter, $\partial C_{Xl}/\partial s_i$. We determine these
derivatives with finite differences, with the power spectra calculated
using the {\sc cmbfast} code \cite{SZ,ZSB}.  For some parameters these
derivatives vary significantly within $1\sigma$ of the target value,
indicating that for these directions of parameter space the likelihood
function is not well fit by a Gaussian. In these cases we use an
iterative method such that the step size for each parameter used to
calculate the derivative is approximately equal to its estimated
$1\sigma$ error. We have found that the resulting Gaussian
approximation gives a good fit to the true likelihood function, in
particular correctly reproducing the strong correlations between some
parameters. However, this dependence of the results on the details of
the algorithm demonstrates that the given parameter errors should be
taken as a guide only. Other uncertainties in the Fisher matrix
approach are considered by Magueijo \& Hobson \shortcite{MH}.

There is also freedom in the choice of the scale
about which to expand the power spectrum, $k_0$ of
Eq.~(\ref{psexp}). The estimated errors on $n$ and its derivatives are
particularly sensitive to the choice of this scale, with values
varying by an order of magnitude for different choices within the
observed range of scales. However, an appropriate choice of $k_0$ can
minimize these error estimates. In fact, when the true power spectrum
is such that only $n$ and $dn/d\ln k$ are non-vanishing it can be
shown\footnote{We are grateful to Daniel Eisenstein for pointing this
out.} that $k_0$ can be chosen so that the error on $n$ is unchanged
from the case of a power-law spectrum, and is uncorrelated with
$dn/d\ln k$. This is because a scale exists at which the
introduction of $dn/d \ln k$ does not contribute to the mean value of
$n$, the mean being weighted by the error estimation procedure.  For
the non-polarized configuration this optimal choice is $k_0=0.064$
Mpc$^{-1}$, while for the polarized configuration it is $k_0=0.045$
Mpc$^{-1}$. Although in principle there is no guarantee that this is
still the case when higher derivatives are present in the power
spectrum, in practice we find that the above choices for $k_0$ give
approximately the same errors on $n$ and $dn/d\ln k$ when $d^2/d(\ln
k)^2$ is included. We therefore give error estimates evaluated at
these scales, bearing in mind that the errors at other scales can be
much larger.

The resulting parameter errors are displayed in
Table~\ref{errors_tab}, for the two different experimental
configurations and for increasingly general forms of the inflationary
perturbation spectrum. For a power-law spectrum our estimates are in
good agreement with those of other workers, and do not change
substantially when the power-law assumption is relaxed, except that as
noted above the errors on $n$ and its derivatives increase
dramatically at scales away from $k_0$. It should be remembered that
these values are not exact one-sigma errors, given the uncertainties
of the Fisher matrix formalism mentioned previously.

\section{Scale-dependent spectral indices in slow-roll inflation}

The results for the spectra from slow-roll inflation are by now extremely 
well known. We follow the definitions of Liddle \& Lyth \shortcite{LL}. The 
spectrum ${\cal 
P}_{\cal R}$ of the curvature perturbation ${\cal R}$ is given by
\begin{equation}
{\cal P}_{\cal R}(k) = \left( \frac{H}{\dot{\phi}} \right)^2 \left(
	\frac{H}{2\pi} \right)^2 = \frac{8}{3 m_{{\rm Pl}}^4} \, 
\frac{V}{\epsilon} \,.
\end{equation}
Here $V(\phi)$ is the potential of the scalar field driving inflation, and 
$\epsilon$ is one of three slow-roll parameters defined by
\begin{eqnarray}
\epsilon & = & \frac{m_{{\rm Pl}}^2}{16 \pi} \left( \frac{V'}{V}
	\right)^2 \,; \\
\eta & = & \frac{m_{{\rm Pl}}^2}{8 \pi} \, \frac{V''}{V} \,; \\
\xi^2 & = & \frac{m_{{\rm Pl}}^4}{64 \pi^2} \,
	\frac{V' V'''}{V^2} \,,
\end{eqnarray}
where a prime indicates derivative with respect to the scalar field $\phi$. 
Note that $\epsilon$ is positive by definition. Despite the square, $\xi^2$ 
can be either positive or negative; the square is to indicate that it is 
second-order in the slow-roll expansion, which is an expansion in $(m_{{\rm 
Pl}} \, d/d\phi)^2$ \cite{LPB}. These parameters must be less than one for 
the 
slow-roll approximation to be valid.

At a given point on the potential, corresponding to the location where the 
expansion in Eq.~(\ref{psexp}) is carried out, the slow-roll parameters are 
in general independent since the derivatives can be freely chosen. There are 
further ones corresponding to higher derivatives but we don't need them in 
this paper.

The equations of motion for inflation give the slow-roll result
\begin{equation}
\frac{d}{d \ln k} \simeq - \frac{m_{{\rm Pl}}^2}{8\pi} \, \frac{V'}{V} \, 
	\frac{d}{d\phi} \,,
\end{equation}
which allows us to compute the spectral index and its derivative with 
respect to the potential. The formulae are remarkably simple; the spectral 
index is \cite{LL92}
\begin{equation}
n = 1 - 6 \epsilon + 2 \eta \,,
\end{equation}
and its derivative is \cite{KT}
\begin{equation}
\frac{dn}{d \ln k} = - 24 \epsilon^2 + 16 \epsilon \eta - 2 \xi^2 \,.
\end{equation}

Along with these, another crucial inflationary observable is the influence 
of gravitational waves, relative to density perturbations, on large-angle 
microwave background anisotropies, given by \cite{LL92} 
\begin{equation}
r_{{\rm ts}} \equiv \frac{C_2({\rm grav})}{C_2({\rm dens})} 
	\simeq 14 \epsilon \,.
\end{equation}

Combining these equations gives
\begin{equation}
\label{scaldep}
\frac{dn}{d \ln k} \simeq 0.12 r_{{\rm ts}}^2 - 0.57 r_{{\rm ts}} (1-n) - 2 
\xi^2 \,,
\end{equation}
which was first given by Kosowsky \& Turner \shortcite{KT}. Remember
that in general $\xi^2$ is independent of the other two parameters,
and that it can be positive or negative. In Fig.~1 we show the
expected scale dependence under the assumption that $\xi^2 = 0$; since
in general it is unconnected to the first two this should be a good
guide though obviously it is possible for some degree of cancellation
to occur. For example, in the power-law inflation model the $\xi^2$
term precisely cancels the scale dependence, though this is a very
special situation.

\begin{figure}
\centering 
\leavevmode\epsfysize=6.3cm \epsfbox{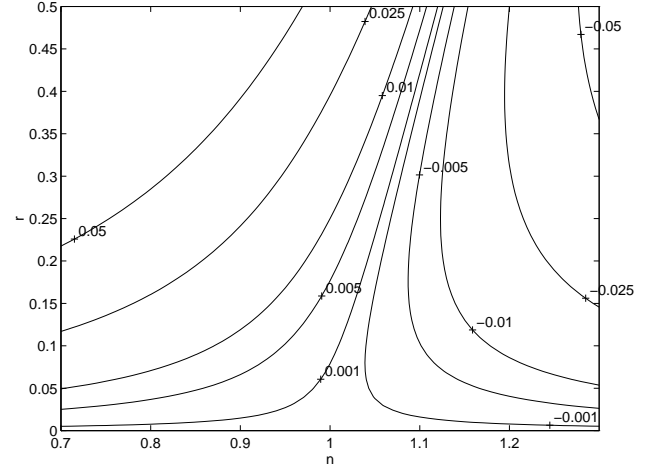}\\ 
\caption[nrcont]{\label{nrcont} This shows contours of constant $dn/d \ln 
k$, assuming $\xi^2 = 0$. Detection of the scale dependence at 95\% 
confidence requires roughly $|dn/d \ln k| \ga 0.01$.}
\end{figure}

As we confirmed in Section \ref{s:ian}, gravitational waves are only
detectable (at 95\% confidence level) if $r_{{\rm ts}} \ga 0.1$ even
if cosmic microwave background polarization is measured \cite{Zetal}. The 
most optimistic
scenario for measuring the scale dependence is if we can assume that
all higher derivatives are negligible, and then with polarization a
95\% confidence detection is possible if $|dn/d \ln k| \ga
0.01$. Studying the first term of Eq.~(\ref{scaldep}), we immediately
see that for it to be important then the tensors have to be
detectable, by some margin. Much more interesting is the second term,
because provided $n$ is not too close to unity, it can give a
detectable contribution to the scale dependence even if $r$ is not
itself detectable (though $r_{{\rm ts}}$ cannot be very small). This is
illustrated in Fig.~1; there exist regions where $|dn/d \ln k| > 0.01$
but where $r_{{\rm ts}} < 0.1$. In fact, hybrid inflation models often
have undetectable $r_{{\rm ts}}$ but $n$ significantly different from
unity. The final $\xi^2$ term in Eq.~(\ref{scaldep}) is in general
independent of the others and so nothing can be said directly, other
than that {\it a priori} it is as likely to reinforce the scale
dependence as partially cancel it.

In conclusion, then, there exist significant parameter regions,
including ones currently observationally viable, where the scale
dependence of the spectral index can be measured. In such cases, it
provides extra information concerning the inflation model which would
not otherwise be available. In some cases, the scale dependence is
measurable even when the tensor contribution is not.

\section{Specific hybrid inflation models}

We end with a brief description of a particular type of inflation
model where $dn/d\ln k$ is detectable. Stewart \shortcite{ewan1,ewan2}
has recently presented two interesting models, the idea being to use
quantum corrections to flatten the potential of the inflaton sufficiently
to allow it to support an inflationary epoch. The argument is not
difficult to reproduce. Typical inflation models require a non-zero
scalar potential energy $V=V_0$. In supergravity this induces soft
supersymmetry-breaking masses for all scalar fields leading to a
classical potential:
\begin{equation}
V=V_0\left[1-{1\over2}A\phi^2 + ...\right]
\label{softsusy} 
\end{equation}
with $|A| \sim 1$. (Here the stringy Planck mass $M_{{\rm Pl}} \equiv
m_{{\rm Pl}}/\sqrt{8\pi}$ has been set equal to one.)  Unfortunately
this does not lead to slow-roll inflation, since $|V''/V| \simeq |A|
\sim 1$. Stewart pointed out that if $\phi$ has either gauge or Yukawa
couplings, to vector or chiral superfields with soft supersymmetry
breaking masses squared of order $V_0$, then quantum corrections will
renormalize the mass of $\phi$ leading to a one-loop renormalization
group effective potential of the form
\begin{equation}
V=V_0\left[1-{1\over2}f(\delta \ln\phi)\phi^2 + ...\right]
\label{effpot} 
\end{equation}
where $\delta \ll 1$ is the one--loop suppression factor, $f(0) = A +
O(\delta)$, with the expression being valid for $V_0 \ll \phi^2 \ll
1$. Such a correction can lead to slow--roll inflation in certain
regions of the potential. In particular, if $\phi_*$ is defined by
\begin{equation}
f_* + {\delta\over 2} f'_* = 0
\label{rengp} 
\end{equation}
where $f_* = f(\delta\ln\phi_*)$ and $f'_*(x) = df_*/dx$ then $|V''/V|
\sim O(\delta)$ in the vicinity around $\phi_*$ and the loop-corrected
potential is flat enough to drive inflation even though the bare
potential is not. The mechanism can work for $f'_*$ either positive or
negative. If it is less than zero then $\phi_*$ is a minimum of the
potential and so a hybrid inflation mechanism is required to end
inflation at some critical value $\phi_{{\rm c}}$; such a model is
constructed in Stewart's second paper \cite{ewan2}. If $f'_* > 0$ then
$\phi_*$ is a maximum of the potential. Stewart \shortcite{ewan1}
considered such a model where $\phi$ rolled towards the true vacuum at
$\phi \sim 1$.

We are interested in the spectral index obtained and its
slope. Introducing $g(y) \equiv -f_*(x) - \delta/2 f'_*(x)$ where $y
\equiv \delta \ln(\phi_*/\phi)$ and defining $\delta$ by
$\delta\ln\phi_* = -1$, Stewart showed that, to lowest order in the
slow-roll approximation, the spectral index $n$ was given by
\begin{eqnarray}
n&=&1 - 2g'_*\delta + 8(1+A)\left(1-{1\over\sqrt{1+3/4(1+A)}}\right) 
	\times 	\nonumber \\  
 & & \hspace*{-0.4cm} \exp\left[ -2+{1\over\sqrt{1+3/4(1+A)}} -g'_*\delta 
 	(N-N_{\rm fr}) \right] \,, 
\label{ewanindex}
\end{eqnarray}
where $g'_* = 2(1+A)/(A+\sqrt{A(A+1)}\,)$, $N-N_{\rm fr}$ is the
number of $e$-folds of inflation from when the scale $k \propto
e^{-N}$ leaves the horizon to $\phi = \phi_{\rm fr}$, the value of
$\phi$ when it begins to fast-roll down the potential and inflation
comes to an end. Since $d/d\ln k \simeq -d/dN$, it follows from
Eq.~(\ref{ewanindex}) that
\begin{eqnarray}
{dn \over d\ln k} &=& 8(1+A)\left(1-{1\over\sqrt{1+3/4(1+A)}}\right) 
	\times \\ 
& & \hspace*{-0.7cm} g'_*\delta \exp\left[ -2+{1\over\sqrt{1+3/4(1+A)}}
	-g'_*\delta (N-N_{{\rm fr}}) \right] \,. \nonumber 
\label{ewanslope}
\end{eqnarray}

Typically the number of $e$-folds of inflation remaining after
observable scales leave the horizon is 40 to 60, depending on the
energy scale and the reheating efficiency. However, this may be
reduced by 10 or 20 if there is a second, low-energy period of
inflation known as thermal inflation \cite{LS}. Low values of
$N-N_{{\rm fr}}$ favour detectable scale dependence.

The most natural scale for such a particle physics motivated potential
is $V_0 \sim M_{{\rm SUSY}}^4$ where $M_{{\rm SUSY}}$ is the
supersymmetry breaking scale in our vacuum. A value $M_{{\rm SUSY}}
\sim 10^{10}$--$10^{11}$ GeV corresponds to a gravity-mediated
supersymmetry breaking. The corresponding COBE normalized value of
$\phi_*$ gives $\phi_* \sim 10^{-11}$, hence $\delta \sim 0.04$, a
number consistent with that derived from a gauge coupling strength
similar to that of the Grand Unified Theory gauge coupling inferred
from LEP data, $\alpha_{{\rm GUT}} \sim 0.04$ \cite{ewan2}. However,
there is significant flexibility in the allowed values of $A$ and
$V_0$, and hence $\delta$. A few values are tabulated in
Table~\ref{dndlnk_tab} to give a feel for the possibilities. What is
encouraging is that provided there is a low number of $e$-foldings,
the scale dependence can be detected by Planck or similar.

\begin{table} 
\caption{\label{dndlnk_tab} Values of $n$ and $dn/d\ln k$ in different 
regions of parameter space.} 
\centering
\begin{tabular}{lllll} 
$A$ & $\delta$ & $e$-folds & $n$ & $dn/d\ln k$ \\ 
\hline 
1 & 0.02 & 20 & 1.32 & 0.013 \\ 
1 & 0.02 & 40 & 1.13 & 0.007 \\ 
1 & 0.04 & 20 & 1.07 & 0.013 \\ 
1 & 0.04 & 40 & 0.92 & 0.003 \\        
2 & 0.02 & 20 & 1.44 & 0.013 \\ 
2 & 0.02 & 40 & 1.23 & 0.008 \\ 
2 & 0.04 & 20 & 1.18 & 0.015 \\ 
2 & 0.04 & 40 & 0.99 & 0.005 \\        
\hline
\end{tabular}
\end{table}

\section{Conclusions}

We have considered the likely magnitude and impact of scale dependence
of the spectral index in inflationary cosmologies. We have estimated
the magnitude necessary to make the scale dependence detectable, and
shown that slow-roll models of inflation, especially those of the
hybrid inflation type, may give an observable effect. If so, this
provides extra information on the inflaton potential which would not
otherwise be available.

If scale dependence of $n$ is considered, then
there is a scale (around $k_0\sim 0.05$ Mpc$^{-1}$) at which the
uncertainty in the determination of $n$ is unchanged from the case of
a power-law spectrum when the first derivative of $n$ is
introduced. We find negligible degradation of the error on $n$ even
when a second derivative is introduced, and have determined the
anticipated errors on the first two derivatives of $n$ at the
preferred scale. We have found that at other scales the errors on $n$
are significantly greater; up to a factor of around 10 greater for
scales corresponding to the current horizon. 

The introduction of extra parameters to model the cosmology implies
that all the cosmological parameters will be more poorly determined.
Fortunately, we have shown that the degradation of the uncertainty in
these parameters is a small effect when the two derivatives are
introduced.

\section*{ACKNOWLEDGMENTS}

EJC is supported by PPARC and IJG and ARL by the
Royal Society. We are particularly grateful to Daniel Eisenstein for
pointing out that the error on the spectral index is not degraded when
its first derivative is included, if evaluated at the appropriate
scale.  We also thank Pedro Ferreira, Wayne Hu, Andrew Jaffe,
Rocky Kolb and Martin White for helpful comments on this work. We
acknowledge use of the Starlink computer system at the University of
Sussex.



\bsp
\end{document}